%% file: main.tex
\documentclass[sigconf, screen, dvipsnames, nonacm]{acmart}

\usepackage{hyperref}
\usepackage[noend]{algorithmic}
\usepackage{algorithm}
\usepackage{multirow}
\usepackage{amsmath}
\usepackage{bbm}
\usepackage{booktabs}
\usepackage[inline]{enumitem} 
\usepackage{subcaption}
\usepackage{bm} 
\usepackage{xcolor}

\usepackage{lipsum}

\usepackage{wrapfig}

\usepackage{arydshln}
\makeatletter
\def\adl@drawiv#1#2#3{
        \hskip.5\tabcolsep
        \xleaders#3{#2.5\@tempdimb #1{1}#2.5\@tempdimb}%
                #2\z@ plus1fil minus1fil\relax
        \hskip.5\tabcolsep}
\newcommand{\cdashlinelr}[1]{%
  \noalign{\vskip\aboverulesep
          \global\let\@dashdrawstore\adl@draw
          \global\let\ adl@draw\adl@drawiv}
  \cdashline{#1}
  \noalign{\global\let\adl@draw\@dashdrawstore
          \vskip\belowrulesep}}
\makeatother

\usepackage{tikz}
\usetikzlibrary{automata, arrows, bayesnet, bending}

\usepackage{tikzscale}

\copyrightyear{2025}
\acmYear{2025}
\setcopyright{rightsretained}

\begin{document}
\title{$t$-Testing the Waters}
\subtitle{Empirically Validating Assumptions for Reliable A/B-Testing}

\author{Olivier Jeunen}
\affiliation{
  \institution{aampe}
  \city{Antwerp}
  \country{Belgium}
}
\email{olivier@aampe.com}

\begin{abstract}
A/B-tests are a cornerstone of experimental design on the web, with wide-ranging applications and use-cases.
The statistical $t$-test comparing differences in means is the most commonly used method for assessing treatment effects, often justified through the Central Limit Theorem (CLT).
The CLT ascertains that, as the sample size grows, the sampling distribution of the Average Treatment Effect converges to normality, making the $t$-test valid for sufficiently large sample sizes.
When outcome measures are skewed or non-normal, quantifying what ``\emph{sufficiently large}'' entails is not straightforward.

To ensure that confidence intervals maintain proper coverage and that $p$-values accurately reflect the false positive rate, it is critical to validate this normality assumption.
We propose a practical method to test this, by analysing repeatedly resampled A/A-tests.
When the normality assumption holds, the resulting $p$-value distribution should be uniform, and this property can be tested using the Kolmogorov-Smirnov test.
This provides an efficient and effective way to empirically assess whether the $t$-test's assumptions are met, and the A/B-test is valid.
We demonstrate our methodology and highlight how it helps to identify scenarios prone to inflated Type-I errors.
Our approach provides a practical framework to ensure and improve the reliability and robustness of A/B-testing practices.
\end{abstract}

\maketitle

\input{1_Intro}
\input{2_Method}
\input{3_Experiments}
\input{4_Conclusions}


\bibliographystyle{ACM-Reference-Format}
\bibliography{bibliography}

\end{document}

%% file: 1_Intro.tex
\section{Introduction \& Motivation}
Contemporary technology companies have embraced continuous experimentation practices to guide their research and product development cycles.
Designing, running, monitoring and subsequently interpreting A/B-tests has become part of the daily routine for many industry practitioners~\cite{kohavi2020trustworthy, Larsen2024}.
The A/B-testing framework is generally well understood, due to a long history of scientific literature on randomised controlled experiments~\cite{Fisher1921,Rubin1974}.
Other recent works highlight pitfalls that often occur in modern-day settings~\cite{Kohavi2022, Dmitriev2017,Jeunen2023_Forum, Kohavi2024}, or focus on ensuring that the metrics under consideration have sufficient statistical power~\cite{Baweja2024,Jeunen2024_Learning,Jeunen2024_RecSysIndustry,Tripuraneni2024, Deng2024, Guo2021}.

When all data has been collected in a properly randomised manner, the next step is typically to estimate Confidence Intervals (CIs) for Average Treatment Effects (ATEs) across a range of user behaviour metrics, and to perform statistical hypothesis tests to ensure observed effects are not simply the result of noise~\cite{Greenland2016}.
The null hypothesis is often a ``\emph{zero effect}'' hypothesis, and hence the procedure involves computing sample means and standard deviations to construct normal CIs and checking whether they contain zero.
More generally, one can construct a test statistic from the CI to inform a $p$-value.
The reason that normal CIs are an appropriate choice is that we are estimating the \emph{average} treatment effect.
According to the CLT, we are guaranteed that as the sample size increases, the sampling distribution of the ATE will converge to normality~\cite{Durret2019, Fischer2010}.
Whilst asymptotic convergence to a normal distribution is clearly a desirable property, practitioners are left with little guidance as to when the sample size is considered ``\emph{sufficiently large}''.
To complicate things further, the underlying distribution of the event under consideration non-trivially impacts the distribution of its mean, and hence, the validity of the outlined approach.

In this work, we describe a practical approach to empirically validate whether we can trust the CIs and resulting $p$-values.
When the null hypothesis holds true by design, we expect a CI with confidence level $100\cdot(1-\alpha)\%$ to \emph{exclude} zero in $\alpha\%$ of cases.
Furthermore, the distribution of $p$-values should be expected to be uniform.
We can emulate such situations by repeatedly resampling synthetic A/A-tests from the available data, leveraging the Kolmogorov-Smirnov test to ascertain whether the resulting $p$-value distribution is likely to be uniform~\cite[p.~188]{kohavi2020trustworthy}.
This provides an efficient and effective approach to empirically validate the assumptions that permeate common practice, allowing us to pinpoint problematic experimental designs where the false positive rate deviates from $\alpha$~\cite{Kohavi2024}.
This, in turn, enforces statistical rigour and increases the trust we can place in A/B-testing estimands that do pass the test.

We note that the statistical hypothesis testing framework and the use of $p$-values itself is a topic of ongoing debate, with several works providing guidance on alternative communication about estimated effects and their associated uncertainty~\cite{Wasserstein2019,McShane2019}.
Other works suggest the use of Bayesian alternatives to handle uncertainty estimation of A/B-testing results~\cite{Deng2015,Gronau2021}.
Either line of work suggests to put more emphasis on confidence (or credible) intervals rather than a $p$-value or binary ``\emph{significant}'' label.
The validity of both frequentist and Bayesian intervals---i.e. whether they fulfil the meaning that is ascribed to them---is a primary concern of our work, and covered by our analysis and methodology.

The following sections formalise the problem statement and our proposed approach, providing empirical results and insights from its implementation in a real-world system.

%% file: 2_Method.tex
\section{Problem Statement \& Methodology}
\subsection{Estimating a confidence interval for the ATE}
Our aim is to leverage online controlled experiments to assess the ATE of an intervention on some outcome of interest $Y$.
Without loss of generality, we assume that this random variable indicates a logged user-level event (e.g. a user opens the app, clicks, converts, renews, churns, et cetera).
We denote the intervention by superscript, for treatment $Y^{\rm T}$ and control $Y^{\rm C}$.
Our estimand is then, with an expectation over experiment randomisation units (i.e. users):
\begin{equation}
    \mathop{\rm ATE}\limits_{{\rm C} \to {\rm T}}(Y) = \mathbb{E}[Y^{\rm T}-Y^{\rm  C}].
\end{equation}

A straightforward estimator for the ATE is given by the difference in sample means.
For a set of users belonging to a \emph{group} (i.e. control C, treatment T, or a general A/B-testing group A) and their observed outcomes $Y$, we have:
\begin{equation}
    \mu_{\rm A}(Y) = \frac{1}{|\mathcal{U}_{\rm A}|} \sum_{i \in \mathcal{U}_{\rm A}} Y_i,\, \, \, \, \enskip \text{and} \, \, \, \, \enskip \widehat{\mathop{\rm ATE}\limits_{{\rm C} \to {\rm T}}}(Y)  =  \mu_{\rm T}(Y) - \mu_{\rm C}(Y).
\end{equation}

To quantify the uncertainty in the estimate, we wish to construct a confidence interval.
The CLT tells us that the distribution of $\mathbb{E}[Y]$ converges to a normal distribution, and hence, so does the distribution for the ATE.
This implies that we can compute:
\begin{align}
    \sigma^{2}_{\rm A}(Y) &= \frac{1}{|\mathcal{U}_{\rm A}|}\sum_{i \in \mathcal{U}_{\rm A}}(Y_i - \mu_{\rm A}(Y))^{2}, \\
   \text{SE}\left(\widehat{\mathop{\rm ATE}\limits_{{\rm C} \to {\rm T}}}(Y)\right) &= \sqrt{\frac{\sigma^{2}_{\rm C}(Y)}{|\mathcal{U}_{\rm C}|} + \frac{\sigma^{2}_{\rm T}(\rm Y)}{|\mathcal{U}_{\rm T}|}}.
\end{align}
A $100\cdot(1-\alpha)\%$ confidence interval can then be obtained as:
\begin{equation}
    \widehat{\mathop{\rm ATE}\limits_{{\rm C} \to {\rm T}}}(Y) \pm  \Phi^{-1}\left(1-\frac{\alpha}{2}\right)\cdot\text{SE}\left(\widehat{\mathop{\rm ATE}\limits_{{\rm C} \to {\rm T}}}(Y)\right),
\end{equation}
where the inverse cumulative distribution function for the standard normal distribution $\Phi^{-1}$ gives the critical value for confidence level $\alpha$.
It should include the ground truth ATE in $100\cdot(1-\alpha)\%$ of cases.
A confidence interval around the ATE is a crucial component to consider when properly interpreting A/B-testing results.

In  a statistical hypothesis testing framework, when zero is not contained by this interval, the null hypothesis is rejected and the result is deemed significant at level $\alpha$.
Alternatively, we can construct a two-tailed $p$-value as:
\begin{equation}\label{eq:pvalue}
    p = 2 \left(1- \Phi\left( \left| \underbrace{\frac{\widehat{\mathop{\rm ATE}\limits_{{\rm C} \to {\rm T}}}(Y)}{\text{SE}\left(\widehat{\mathop{\rm ATE}\limits_{{\rm C} \to {\rm T}}}(Y)\right)}}_{z\text{-score}} \right| \right)\right),
\end{equation}
and reject the null hypothesis when $p < \alpha$.
The $p$-value can be described as the probability of observing results at least as extreme as what is observed, given that the null hypothesis holds true.

The meaning that is ascribed to both confidence intervals and $p$-values relies heavily on the assumption that the distribution of the estimand has approached normality. 
Whilst the CLT guarantees this property to hold asymptotically, finite sample scenarios require us to empirically validate that the above procedure is appropriate. 

\subsection{Empirically validating confidence intervals}
Naturally, directly validating whether the obtained CI includes the true ATE would require knowledge of the latter, which is prohibitive.
Alternatively, A/A-tests allow us to emulate experiments where we know the true ATE by design (i.e. $0$, as the null hypothesis holds).
This enables us to estimate the empirical coverage of the obtained CIs, through repeated resampling of A/A-groups.
When the distribution of the ATE has approached normality, the distribution of the $p$-values that we obtain over resampled A/A-tests should resemble a uniform distribution.
The Kolmogorov-Smirnov test provides a rigorous statistical framework to flag cases where it does not.
This allows us to, for a set of users $\mathcal{U}$ and outcomes $Y$, assess how amenable the data is to reliable estimation of CIs on ATE($Y$) using the above-mentioned standard methods.

As such, we repeatedly resample groups ${\rm A}_{i}$,${\rm A}_{i}^\prime$ for $n$ iterations, obtain $n$ confidence intervals for $\widehat{\mathop{\rm ATE}\limits_{{\rm A}_i \to {\rm A}_{i}^{\prime}}}(Y)$ and obtain a set of $p$-values $\{p_{1},\ldots,p_{n}\}$.
Given the empirical Cumulative Distribution Function (eCDF) of $p$-values $F_{\rm emp}(p)$, we wish to assess how it deviates from the uniform distribution with CDF $F_{\rm uni}(p)=p$ for $p\in[0,1]$.
The test statistic leveraged by Kolmogorov-Smirnov, known as the $D$-statistic, measures the $\infty$-norm over the observed differences between the two CDFs as:
\begin{equation}
    D = \sup_{p\in[0,1]} \left|F_{\rm emp}(p) - F_{\rm uni}(p)\right|.
\end{equation}
Under the null hypothesis that the distributions are equivalent, $D$ follows a Kolmogorov distribution.
As such, we can obtain a $p$-value that is used to reject the null hypothesis that the $p$-values obtained from the A/A-tests are uniformly distributed, or, that we have sufficient samples to reliably estimate treatment effects on $Y$.
Note that this is equivalent to testing whether the $z$-scores in Equation~\ref{eq:pvalue} follow a standard normal distribution.
Arguments against the statistical hypothesis testing framework apply here as well.
We suggest to pay special attention to cases where the $D$-statistic is high, or conversely, the Kolmogorov Smirnov $p$-value is low, rather than assigning binary (non-)significant labels to metrics.

In cases where an outcome $Y$ is flagged through this procedure, CIs and $p$-values obtained through the $t$- or $z$-test should be considered unreliable.
We can rely on alternative non-parametric methods to estimate CIs on ATE($Y$) in those situations, e.g. based on permutation sampling or bootstrapping.
We note that other commonly used methods like the Mann-Whitney U-test formulate a different null hypothesis, and cannot be used as a drop-in replacement without careful consideration~\cite{Fay2010}.
Furthermore, as the above-mentioned alternatives typically come with a considerable computational cost and specialised engineering solutions, they are significantly less desirable as a default approach.
A deeper exploration of their applicability falls outside of the scope for this work, but provides an interesting avenue for future research.

%% file: 3_Experiments.tex
\section{Experimental Results \& Discussion}
Until now, we have discussed the theoretical aspects and assumptions of treatment effect estimation and uncertainty quantification in the context of A/B-testing.
To assess the practical utility of the aforementioned methods and their potential, we wish to empirically answer the following research questions:

\begin{description}
    \item[\textbf{RQ1}] \textit{Can the Kolmogorov-Smirnov test on resampled A/A-tests uncover outcomes $Y$ for which normal CIs are not appropriate?}
    \item[\textbf{RQ2}] \textit{Does the $D$-statistic provide additional information as a diagnostic over the number of observations alone?}
    \item[\textbf{RQ3}] \textit{Can we leverage other information about the distribution of $Y$?}
\end{description}

To provide empirical answers to these questions, we resort to a subet of a proprietary log of user activity data on a consumer-facing application. 
As our discussions and insights are general and agnostic to the use-cases, we expect our findings to translate to other applications.
The data consists of $|\mathcal{U}|\approx2$  million users, and approximately 17 million logged instances across 50 event types.

\begin{figure}[!t]
    \centering
    \includegraphics[width=\linewidth]{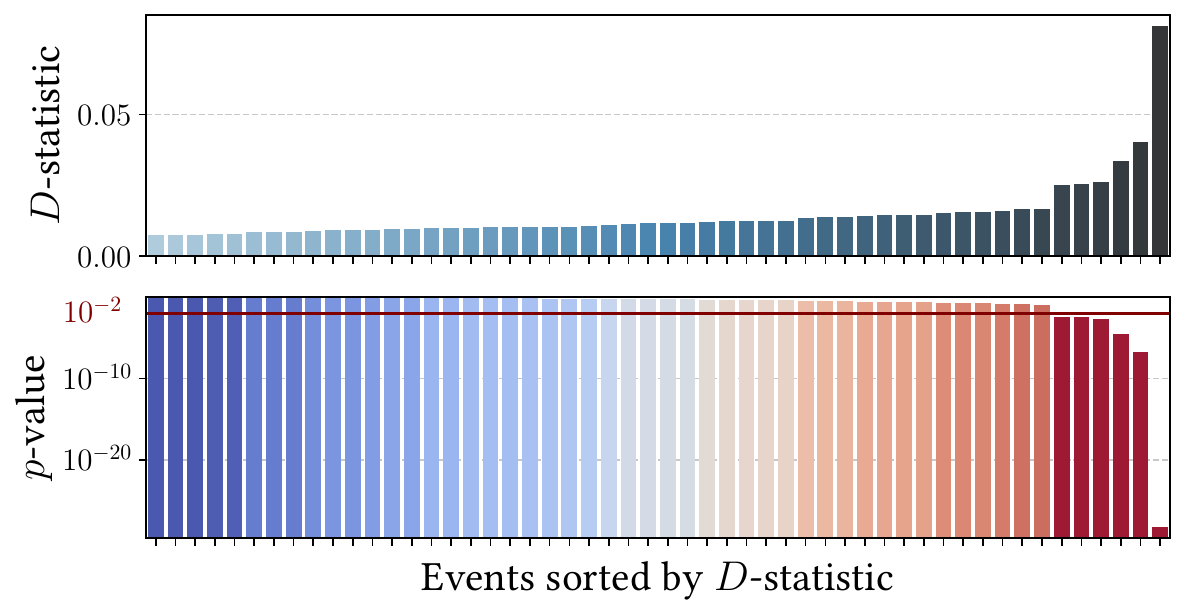}
    \caption{Visualising the Kolmogorov-Smirnov $D$-statistic and resulting $p$-value per user-event we measure. Whilst the majority of $p$-value distributions cannot be distinguished from uniform, we reject the null hypothesis for several.}
    \label{fig:1}
\end{figure}

\paragraph{\textbf{RQ1}: Utility of the approach.}
For every possible outcome $Y$ to measure, we construct $n=5\,000$ synthetic A/A comparisons and collect $D$-statistics and Kolmogorov-Smirnov $p$-values w.r.t. the expected uniform distribution.
Figure~\ref{fig:1} visualises their distribution over event types.
As expected, the null hypothesis that the CLT has sufficiently kicked in cannot be refuted for the majority of outcomes $Y$, and normal CIs are appropriate to reflect the uncertainty in the ATE estimate.
Nevertheless, the procedure succeeds in highlighting several events that require further investigation.\footnote{Note that a direct interpretation of Kolmogorov-Smirnov $p$-values would require a multiple testing correction to be applied~\cite{Shaffer1995}. Even with a crude Bonferroni correction, the ATE($Y$) distribution of several events still violates normality significantly.}

\begin{figure}[!t]
    \centering
    \includegraphics[width=\linewidth]{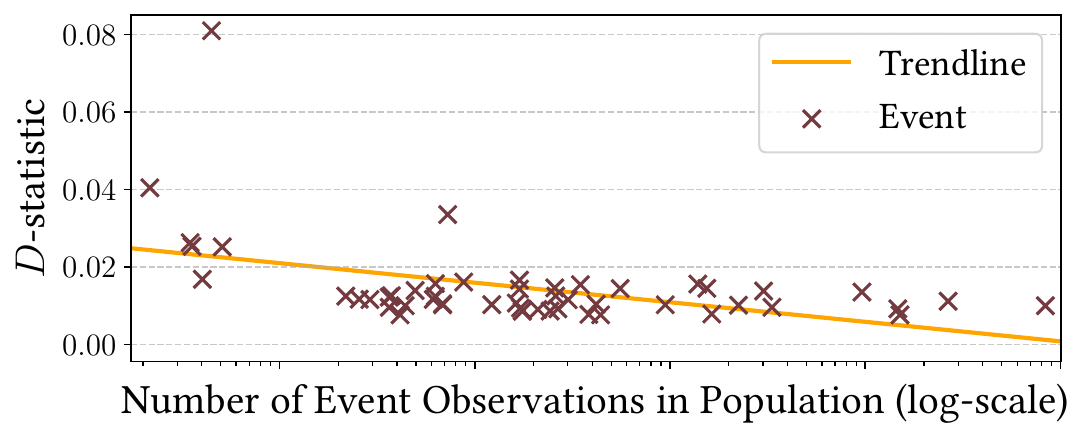}
    \caption{Visualising the number of event observations overall to their $D$-statistic, with a log-linear trendline. Whilst rare events lead to an increase in distribution divergence, the relationship is not monotonic (Spearman's $\rho\approx0.45$).}
    \label{fig:2}
\end{figure}

\paragraph{\textbf{RQ2}: Considering event frequency.}
A natural question to consider is whether the sample size is the deciding factor in determining whether the sampling distribution of the ATE has approached normality sufficiently well enough for the $t$-test to be valid.
Since all comparisons use the full dataset (i.e. roughly 1 million users per A/A group), and the sample size is thus constant, this is clearly not the case.
Instead, we might then consider event frequency, as for rare events the majority of users will not contribute to the ATE.
Figure~\ref{fig:2} visualises the number of event observations on a logarithmic scale, ranging from approximately 200 to 8 million, in relation to the $D$-statistic on the y-axis. 
Whilst a clear correlation is visible (Spearman's $\rho\approx0.45$), there is no monotonic relationship.
This suggests that while event frequency is informative, the $D$-statistic brings additional diagnostic value when assessing $t$-test validity.

\begin{figure}[!t]
    \centering
    \includegraphics[width=\linewidth]{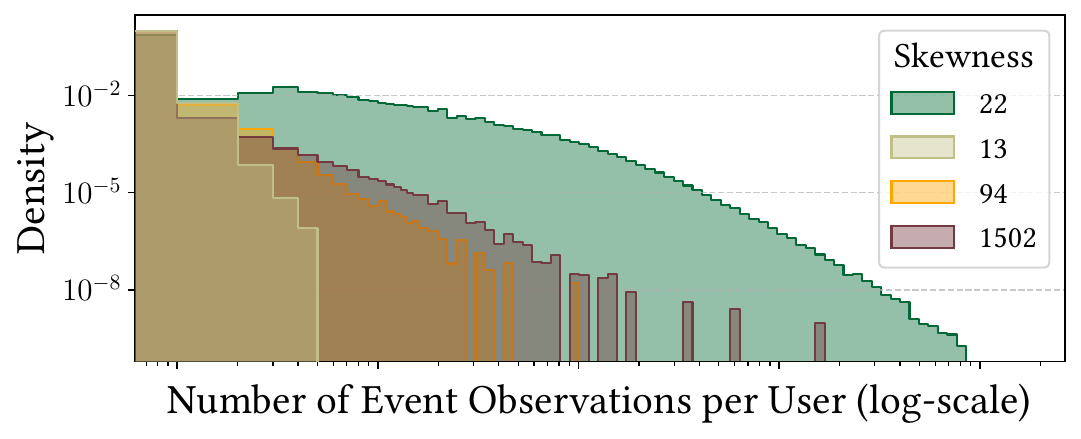}
    \caption{The empirical density function for various events intuitively shows that the sample skewness of the empirical event distribution per user is an indicator of the required sample size for the CLT to kick in, and the mean event distribution to approach normality (Spearman's $\rho\approx0.43$).}
    \label{fig:3}
\end{figure}

\paragraph{\textbf{RQ3}: Exploring other summary statistics for $\mathsf{P}(Y)$.}
Aside from event frequency (i.e. $\mathbb{E}[Y]$), we might be interested in other moments of the outcome distribution $\mathsf{P}(Y)$.
\citet{Kohavi2014} and \cite{Kohavi2022} discuss the skewness as an important diagnostic for CLT appropriateness.
We consider the sample skewness of $Y$, and report it alongside histograms for four events in Figure~\ref{fig:3}.
We visualise, in order of the legend:
\begin{enumerate*}[label=(\roman*)]
  \item the most frequent event,
  \item the rarest event for which normality cannot be rejected,
  \item an event with similar frequency to (iv) but low $D$-statistic, and
  \item the most frequent event with $p$-value < $1e-4$.
\end{enumerate*}
Sample skewness estimates are shown in the legend, suggesting that a higher skewness implies slower CLT convergence.
Nevertheless, there is no monotonic relationship: Spearman's $\rho\approx0.43$ suggests a significant rank-correlation with the $D$-statistic, but confirms the independent informational value of the Kolmogorov-Smirnov test as a diagnostic tool.

%% file: 4_Conclusions.tex
\section{Conclusions \& Outlook}
A/B-tests are omnipresent in modern technology companies, often seen as the ``gold standard'' of experimentation practices.
The default estimand is typically the ATE on a metric of interest, and statistical uncertainty on the estimate is handled through standard methods. 
It is often forgotten that these methods rely on implicit assumptions that might be violated, invalidating the estimates.

Most commonly, we assume that the distribution of the ATE has converged to normality due to the CLT. 
If this is false, the CIs and $p$-values we use to assess A/B-testing outcomes become misleading.
In this work, we propose an efficient and effective manner to validate this assumption empirically, through the use of repeatedly resampled A/A-tests and a Kolmogorov-Smirnov test on the uniformity of the resulting $p$-value distribution.
We describe our approach, present empirical results on real-world data that both elucidate the methods and highlight that potential proxies (like sample size or skewness) are promising but imperfect.
This provides a practical framework to assess A/B-testing estimands and avoid situations where improper estimation methods are applied, in the hope that it will help the community to enforce statistical rigour and avoid inflated false positive risk going forward.